\documentclass[%
 reprint,
 amsmath,amssymb,
 aps,
]{revtex4-1}

\usepackage{graphicx}
\usepackage{dcolumn}
\usepackage{bm}
\usepackage{epstopdf}
\usepackage[mathlines]{lineno}
\usepackage[breaklinks=true,colorlinks=true,linkcolor=blue,urlcolor=blue,citecolor=blue]{hyperref}
\usepackage{array}
\usepackage[table]{xcolor}

\newcolumntype{C}{>{\centering\arraybackslash}m{2cm}}
\begin{document}

\preprint{APS/123-QED}



\title{Generative Agent-Based Models for Complex Systems Research: a review}

\author{Yikang Lu$^{1,\ddag}$}
\author{Alberto Aleta$^{4,5,\ddag}$}
\author{Chunpeng Du$^{3}$}
\author{Lei Shi$^{1,2}$}
\email{shi\_lei65@hotmail.com}
\author{Yamir Moreno$^{4,5,6}$}
\email{yamir.moreno@gmail.com}

\address{$^1$ School of Statistics and Mathematics, Yunnan University of Finance and Economics, Kunming, 650221, China}
\address{$^2$ School of Statistics and Mathematics, Shanghai Lixin University of Accounting and
	Finance, Shanghai, 201209, China}
\address{$^3$ School of Mathematics, Kunming University, Kunming, Yunnan 650214, China}
\address{$^4$ Institute for Biocomputation and Physics of Complex Systems, University of Zaragoza, Zaragoza, 50018, Spain}
\address{$^5$ Department of Theoretical Physics, University of Zaragoza, Zaragoza, 50009, Spain}
\address{$^6$ Centai Institute, Turin, Italy}

\thanks{These authors contributed equally to this work.}
\date{\today}

\begin{abstract}
The advent of Large Language Models (LLMs) has significantly transformed the fields of natural and social sciences. Generative Agent-Based Models (GABMs), which utilize large language models in place of real subjects, are gaining increasing public attention.
Far from aiming for comprehensiveness, this paper aims to offer readers an opportunity to understand how large language models are disrupting complex systems research and behavioral sciences. In particular, we evaluate recent advancements in various domains within complex systems, encompassing network science, evolutionary game theory, social dynamics, and epidemic propagation. Additionally, we propose possible directions for future research to further advance these fields.
\end{abstract}

\keywords{Large Language Models, Complex Network, Game Theory, Nonlinear Dynamic, Generative Agent-Based Models}                       
\maketitle

\section{\label{sec:level1}Introduction}

The emergence of Generative Artificial Intelligence (GenAI), which refers to generative models that can generate text, images, videos, or other types of data, has transformed perceptions within the field of artificial intelligence \cite{jo2023promise,walters2020assessing,liu2021density,noy2023experimental,wang2023scientific}. These models are continually evolving and improving, although they have grown particularly after the introduction of transformed-based neural networks \cite{Vaswani2017}. These models include Large Language Models (LLMs) such as GPT-4 by OpenAI, LLaMA by Meta, Copilot by Microsoft, Gemini by Google, and Ernie by Baidu, among others. But they also encompass image generation models such as DALL·E 3 by OpenAI, Stable Diffusion by Stability AI, or Midjourney by Midjourney, Inc. \cite{pastor2023generative}.

In the realm of complex systems research, we are particularly interested in LLMs. LLMs began to emerge in 2018 and became ubiquitous by the end of 2022. As these models continued to advance, the number of their parameters grew significantly, with GPT-4, for instance, reportedly boasting over 1 trillion parameters \cite{Stern2023Apr,bubeck2023sparks}.  Moreover, these models exhibit promising potential for a variety of scientific applications, showcasing their proficiency in tackling complex problem-solving and knowledge integration tasks. In fact, they have already had a measurable effect on academics' writing \cite{geng2023} and may have a profound impact on the advancement of both social and natural sciences \cite{karinshak2023working,grossmann2023ai,epstein2023art,birhane2023science,de2023chatgpt,noy2023experimental,park202social,ai4science2023impact}. 

In the field of natural sciences, researchers are exploring strategies to reduce the training cost of these models such as using mixture-of-experts architectures \cite{Du2021Dec}, continuously pre-training \cite{ibrahim2024simple}, or using scaling laws to extrapolate during training \cite{gadre2024language}. Other researches focus on optimizing the cost of using LLMs \cite{Shekhar2024Jan,song2023powerinfer} or on mitigating their ecological impact \cite{Stojkovic2024Mar,Rillig2023Mar}. There are also intensive efforts devoted to extending their generative capacities beyond text such as VideoPoet for video generation \cite{kondratyuk2023videopoet}, or AppAgent for creating agents capable of operating smartphone applications \cite{yang2023appagent}.

However, the interest of LLMs in research goes beyond these foundational aspects. In the field of social sciences, LLMs find applications in various domains \cite{bai2023there}. In linguistics, they are utilized for language prediction tasks \cite{kambhampati2024can}. In economics and social sciences, researchers are striving to imbue LLMs with unique personalities, enabling them to operate as individuals to generate synthetic data \cite{cheng2023compost,veselovsky2023generating}. In consumer behavior research, LLMs behavior aligns with economic theory across several dimensions, including downward-sloping demand curves, diminishing marginal utility of income, and state dependence \cite{brand2023using}. In addition, LLMs' decisions in budget allocation scenarios received higher rationality scores than those made by humans \cite{chen2023emergence}. This alignment underscores its capacity to produce authentic survey responses relevant to consumer behavior \cite{brand2023using}. 

In psychological experiments, LLMs' behaviors demonstrated a high degree of congruence with prevailing societal values \cite{dillion2023can}. In multiple-choice question tests, it was shown that LLMs could successfully best the majority baseline and even infer the question given the options \cite{balepur2024artifacts}. In the exploration of fairness and framing effects in sociology, researchers have integrated LLMs as computational models of humans into classic game experiments \cite{horton2023large}.
Similarly, LLMs can replicate the ``wisdom of the crowd'' effect, akin to human behavior \cite{schoenegger2024wisdom}. By replicating human behavioral experiments, these studies reveal both dissimilarities and similarities between human behavior and that exhibited by LLMs, which can be used to study human behavior or to design better surveys and experiments much faster and for a fraction of the cost \cite{argyle2023out,aher2023using}. The demonstrated consistency with human behavior \cite{ren2024emergence,chiang2024chatbot,dillion2023can,ouyang2022training,wu2023q} suggests that LLMs can perform some of the same operations as humans \cite{Huang2023Oct}, thus attracting significant attention from scholars. In particular, Argyle \emph{et al.} provides a very nice overview of how LLMs can be used as effective proxies for specific human subpopulations in social science research \cite{argyle2023out}.

This surge in research content pertaining to LLMs has prompted several review articles from diverse perspectives \cite{xu2024knowledge,bubeck2023sparks,zhang2023survey,xu2024ai,gao2023large}. In this paper, we offer a comprehensive overview of current research in the context of complex systems, with particular emphasis on four areas: (i) complex networks; (ii) game theory from a behavioral perspective; (iii) social dynamics; and, (iv) epidemic modeling. Throughout the paper, we will also discuss the emergence of a novel framework to study complex systems - Generative Agent-Based Models (GABMs) \cite{Ghaffarzadegan2023Sep}. 

\begin{figure}
\centering
\includegraphics[width=0.5\textwidth]{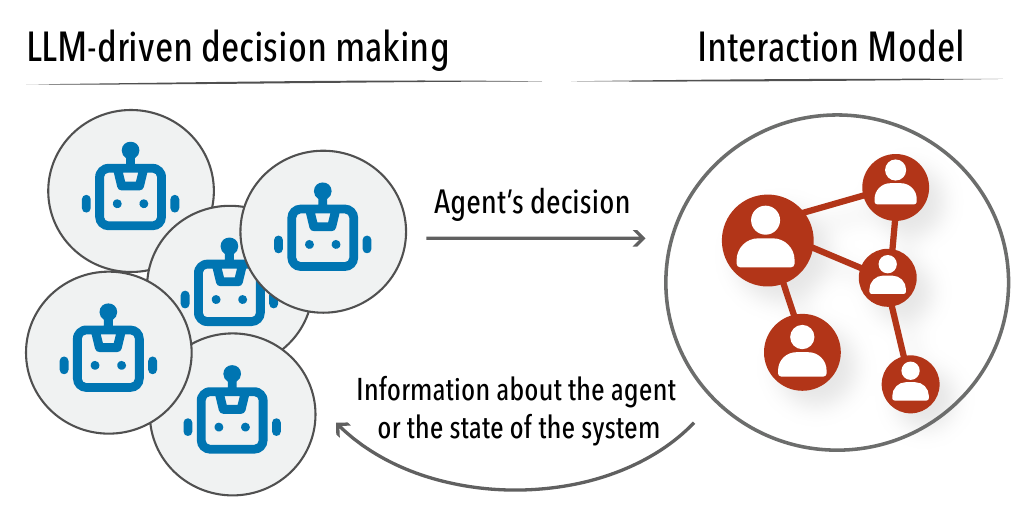}
\caption{\textbf{Generative Agent-Based Models (GABMs)}. In GABMs, agents do not make decisions about their interactions based on a fixed set of rules. Instead, a prompt is sent to an LLM including the desired details and it returns the decision that the agent should follow \cite{Ghaffarzadegan2023Sep}.}
\label{fig:intro}       
\end{figure}

The main idea behind GABMs is that the rules that agents have to follow are not completely fixed a priori. Instead, the decisions of the agent are driven by an LLM whose behavior can be enriched by prompting it with specific information about the problem, the social characteristics of the agent it should represent, or any other feature that is important for the model, as represented in Figure \ref{fig:intro}. For instance, Zhu \emph{et al.} \cite{Zhu2023May} created agents that can play the video game Minecraft using the logic and common sense capabilities of LLMs. This is a good example of how LLM agents can perform advanced operations in complex environments with a success rate much higher than traditional reinforcement-learning controllers.

The paper is structured as follows: First, we discuss relevant works on the use of LLMs to study complex networks in Sec.~\ref{sec:level2}. Next, in Sec.~\ref{sec:level3}, we discuss experiments of cooperative behavior in which LLMs are introduced. Then, in Sec.~\ref{sec:level4} and Sec.~\ref{sec:level5}, we look into various social dynamics and epidemic models coupled with LLMs. Finally, in Sec.~\ref{sec:level6} we provide our summary and perspectives.

\section{\label{sec:level2} Complex networks in the LLMs environment}

Complex networks are one of the fundamental tools in the study of complex systems as they provide a straightforward way to capture the interaction between their constituent elements. For systems in many different domains, these networks share similar properties such as heterogeneous degree distributions or the small-world feature \cite{barabasi2003scale,wang2003complex}. Moreover, in the particular case of human social networks, it has been observed that many nodes are not more than six connections away from any other, something also known as ultrasmall-organization and shown to emerge from human cooperation and altruism \cite{samoylenko2023there}. Thus, the study of these networks can provide answers to interesting questions on human behavior. 

\begin{figure}
\centering
\includegraphics[width=0.5\textwidth]{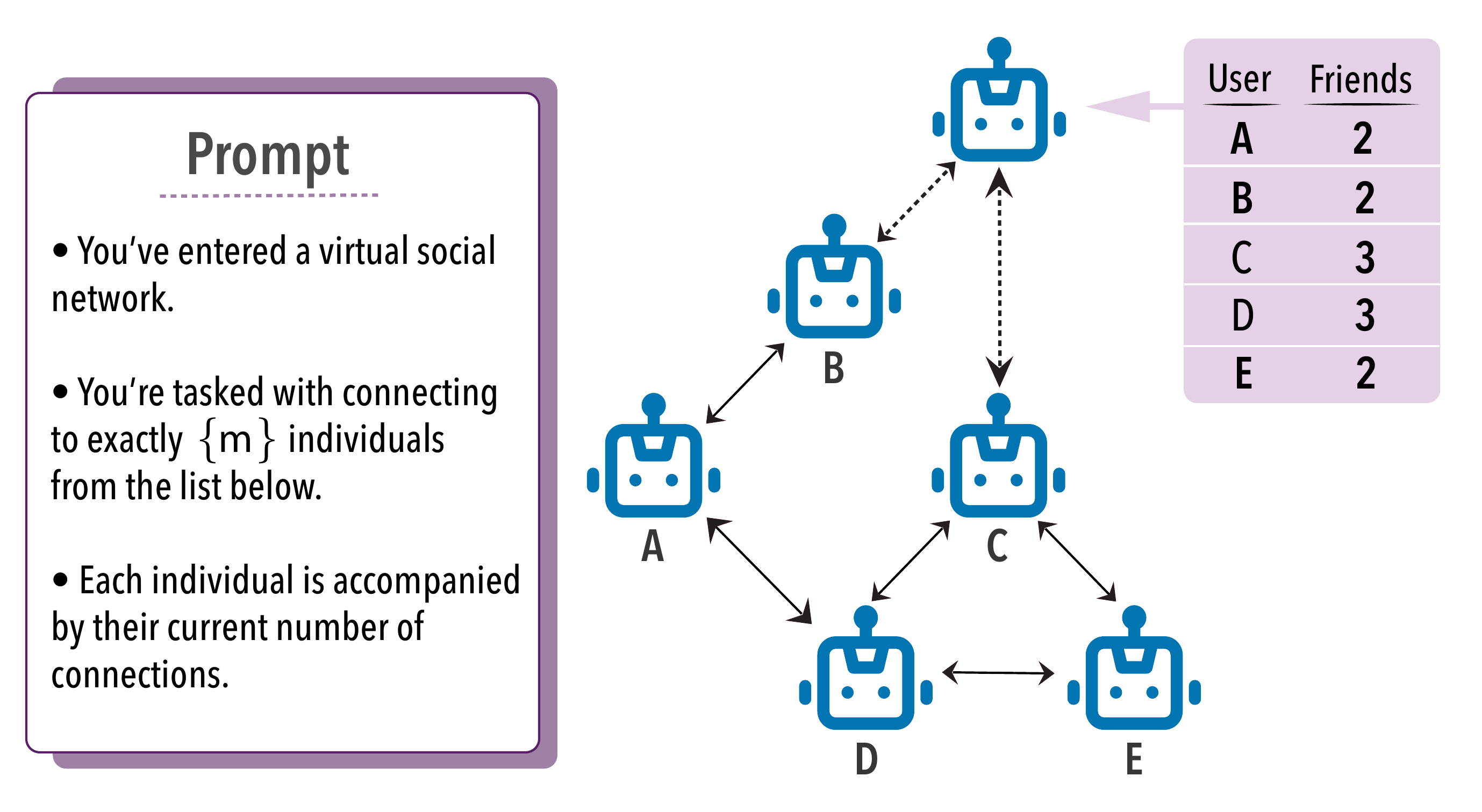}
\caption{\textbf{LLM-based network growth with generative agents}. Each generative agent of a hypothetical online social network is initialized with the prompt shown on the left, together with a comprehensive list of all network users along with their respective number of friends. Then, at each time step, a new agent is added, provided the information, and allowed to establish connections with $m$ other nodes. This iterative process continues until the network reaches the desired size\cite{de2023emergence}.}
\label{fig:network_growth}       
\end{figure}

LLMs can be used to analyze these networks \cite{jiang2023social,mao2024identify}. However, the appearance of artificial agents driven by LLMs opens new questions on the structure and dynamics of the interaction networks studied so far since it is becoming possible to also incorporate human-computer and computer-computer interactions. Along this direction, Park \emph{et al.} \cite{park2023generative} looked into the potential of integrating artificial agents in interactive applications by creating a sandbox environment imitating a small town in which several agents controlled by ChatGPT were given unique personalities and could interact and have human-like daily routines. In their simulation, the agents exhibited seemingly emergent social behaviors, such as when one of them celebrated a party and they started to send invitations to each other. 

To better understand how LLMs may interact, Giordano De Marzo \emph{et al.} \cite{de2023emergence} explored the self-organization of generative agents in forming complex network structures. In their study, nodes represented generative agents whose behavior was controlled by GPT-3.5-turbo. The agents were initialized using the prompts shown in Figure \ref{fig:network_growth}, simulating the growth of an online social network. Upon being initialized, each agent was identified by a randomly assigned 3-character string, and the degree of all other agents was known. Then, they selected their connections following the prompt and established undirected links accordingly, thereby updating the network's degree list. This iterative process persisted, adding new nodes to the network until the desired network size was attained, see Figure \ref{fig:network_growth}.

Following this process, the LLM created a network with a hub-and-spoke structure, that does not resemble the classical results obtained from preferential attachment algorithms \cite{Albert2002Jan}. Interestingly, the researchers found that this was a consequence of a bias in the selection of nodes by the LLM that depended on their name, as seen in \cite{de2023emergence}. By randomizing the names of the agents at each step, they were able to remove this bias, obtaining network structures much closer to the ones obtained with preferential attachment algorithms based on the degree of the nodes. Note that this strategy was not explicitly asked to the LLM, which means that it somehow captured the dynamics of human behavior in social networks. However, it also added an unforeseen bias that had to be corrected to obtain the desired results. Thus, this study highlights some of the limitations of these models and the importance of benchmarking them in a broad set of tasks before they can be used for human behavior research.

Similarly, Lai \emph{et al.} \cite{lai2024evolving} deployed 10 artificial agents based on the LLM Claude-2.1 and allowed them to freely interact without specific priors on what they should do. In particular, they simulated a ``cocktail party'' consisting of 30 communication rounds. In each round, agents who wanted to interact with another one had to send an invitation. If the receiver accepted, they would have a pairwise conversation until either of them decided to end the conversation. The agents showed a tendency to interact repeatedly with the same peers rather than exploring new connections. Furthermore, an analysis of their conversations indicated a certain amount of homophily, another common characteristic of human social networks.

The studies mentioned above provide notable examples of networks formed among artificial intelligencies ruled by the same LLM. However, the networks that might form among intelligences and humans, or the interaction among multiple LLMs have not been comprehensively explored. Besides, there are still many challenges in the implementation of these systems, such as the biases introduced by prompting the agents initially, their training process, or even the restrictions that are usually imposed on them, such as their bias towards positivity \cite{Sharma2023Oct, Mesko2023Jul}.

\section{\label{sec:level3} Game theory in the LLMs environment}

Game theory, as a mathematical framework, offers tools for analyzing and predicting the behavior of rational agents within contexts characterized by uncertainty \cite{roughgarden2010algorithmic}. In recent decades, scholars have extensively examined the inherent factors influencing cooperative behaviors and the mechanisms that foster cooperation, primarily through the lenses of evolutionary games \cite{xia2023reputation,civilini2021evolutionary,zhu2021information,szolnoki2015conformity,pi2022evolutionary,lu2019impacts} and behavioral sciences \cite{perc2019social,jusup2022social}. In traditional evolutionary game models, the evolution of the strategies of the individuals, such as Fermi updating \cite{xia2023reputation,civilini2021evolutionary,zhu2021information}, conformist updating \cite{szolnoki2015conformity,pi2022evolutionary,lu2019impacts}, or self-reversing rules \cite{xiong2021cooperative,quan2022keeping}, must be pre-established.

This paradigm persists even in studies involving basic human-computer interactions like simple bots \cite{szolnoki2016zealots,cardillo2020critical,masuda2012evolution,veselovsky2023generating} and interactions with the environment driven by reinforcement learning-based intelligences \cite{du2024evolution,geng2022reinforcement,lu2023reinforcement}. In addition, numerous experimental studies have investigated real-life gaming behavior to explore the mechanisms underlying the persistence of inter-individual cooperation \cite{wang2018exploiting,shi2020freedom}. Yet, games involving humans and computers, or computers versus computers, have been relatively neglected. While some relevant literature on human-robot interaction exists, the robots discussed in this literature still rely on predefined rules to operate \cite{crandall2018cooperating,makovi2023trust,karpus2021algorithm}. Nonetheless, there are already interesting results. For instance, it was shown that adding some bots to cooperative experiments could increase the cooperation of humans, but also that humans were more likely to exploit AI agents and feel less guilty than when playing with humans \cite{santurkar2023whose}.

In this context, LLMs offer new opportunities thanks to the possibility of creating open-ended agents. This way, the strategies or opinions reflected by the LLMs become a field of study on its own and a new way of exploring human interaction \cite{qian2023communicative}. Along these lines, we can identify, at least, four advantages to using LLMs instead of human participants in evolutionary game experiments \cite{guo2023gpt}. Firstly, assessing the capacity of LLMs to engage in gameplay akin to human performance holds intrinsic value. Secondly, experiments involving LLMs are characterized by lower costs compared to those involving human subjects, as delineated by Horton \cite{horton2023large}, thus facilitating enhanced control over experimental variables, the evaluation of various treatments, and bolstered reproducibility and scalability. Thirdly, such experimentation mitigates certain ethical quandaries typically associated with real-world experiments. Lastly, leveraging the language-based proficiencies of Artificial Intelligences (AIs) may prove advantageous for research endeavors about communication-related subjects \cite{suzuki2024evolutionary}.

For these reasons, in the past couple of years, the research on LLMs applied to behavioral experiments has boomed. We can find studies using artificial agents controlled by LLMs in many different games, including dictator games \cite{aher2023using,horton2023large,guo2023gpt}, rock-paper-scissors games \cite{fan2023can}, prisoner's dilemmas \cite{guo2023gpt,brookins2023playing,Fontana2024Jun}, public goods \cite{lai2024evolving}, and others (see Table \ref{tab1}). LLM-driven agents can mimic intricate internal features of human cognition, and allow researchers to expand their analyses through techniques such as cueing, contextual learning, or fine-tuning, unlike traditional bots \cite{horton2023large}. These experiments illustrate the feasibility of simulating individuals with a wide range of characteristics and traits \cite{suzuki2024evolutionary,capraro2024language}. In the following, we review some of the results obtained in these games.

\begin{table}[ht]
	\centering
	\begin{tabular}{|p{3.75cm}|p{4.75cm}|}
		\hline
		\rowcolor{gray!30}
		\cellcolor{blue!25}Game & \cellcolor{blue!25}Paper \\
		\hline
		\cellcolor{pink!25} The Dictator Game & \cellcolor{pink!25} \cite{fan2023can,chan2023towards,chan2023towards,capraro2024language,horton2023large,sreedhar2024simulating,brookins2023playing,Johnson2023Jan,Xie2024Feb,mei2024turing,mccannon2024artificial,mozikov2024good,babin2024chatbot}\\
		\cellcolor{pink!25} The Ultimatum Game & \cellcolor{pink!25} \cite{chan2023towards,aher2023using,guo2023gpt,sreedhar2024simulating,mei2024turing,mozikov2024good,henry2024prompting}\\
		\cellcolor{pink!25} The Prisoner's Dilemma & \cellcolor{pink!25}\cite{lore2023strategic,chan2023towards,xu2023magic,chan2023towards,guo2023gpt,brookins2023playing,akata2023playing,phelps2023investigating,heydari2023strategic,Fontana2024Jun,Duan2024Feb,mei2024turing,mozikov2024good,herr2024large,roberts2024large} \\
		\cellcolor{pink!25} Public goods& \cellcolor{pink!25}\cite{li2023beyond,xu2023magic,huang2024far,lai2024evolving,mei2024turing,babin2024chatbot} \\
		\hline
	\end{tabular}
	\caption{List of papers with behavioral experiments that include artificial agents driven by LLMs.}
	\label{tab1}
\end{table}

The Ultimatum Game is a non-zero-sum game involving two participants. In this game, one participant acts as the proposer, suggesting a distribution of resources to the other participant, who acts as the responder. If the responder accepts the proposed distribution, the resources are allocated accordingly. However, if the responder rejects the proposal, neither participant receives anything \cite{oosterbeek2004cultural}. 
Aher \emph{et al.} \cite{aher2023using} simulated several economic, psycholinguistic, and social psychology experiments using multiple LLMs (such as DaVinci-002, GPT-3.5 or GPT-4). In the case of the Ultimatum Game, they found that the answers given by the LLMs agree closely with human decision trends. However, they also observed that LLMs were affected by the name and gender of the artificial agents. For instance, agents with male titles were more likely to accept an unfair offer from an agent with a female title. And vice versa, female agents were less inclined to accept an unfair offer from a male agent.
 
One characteristic of this game is that individuals often opt to ``punish'' other players to uphold social norms rather than solely pursuing personal payoffs. Sreedhar \emph{et al.} \cite{sreedhar2024simulating} investigated whether LLMs could replicate this nuanced behavior in a simulation using GPT-3.5 and GPT-4. They compared two architectures: a single-agent LLM, in which the same LLM agent acts as participant and responder; and a multi-agent LLM, in which there are two independent agents. Furthermore, they evaluated their abilities to (1) simulate human-like behavior in an Ultimatum Game, (2) model the personalities of players with traits such as greed and fairness, and (3) develop logically coherent and personality-consistent robust strategies. Their results demonstrated that the multi-agent LLM behavior was consistent with human behavior 88\% of the time, while the single-agent was consistent only 50\% of the time. The major issue in both settings was that the strategies followed by the LLMs were inconsistent with their personality.

\begin{figure*}
	\centering
	\includegraphics[width=0.9\textwidth]{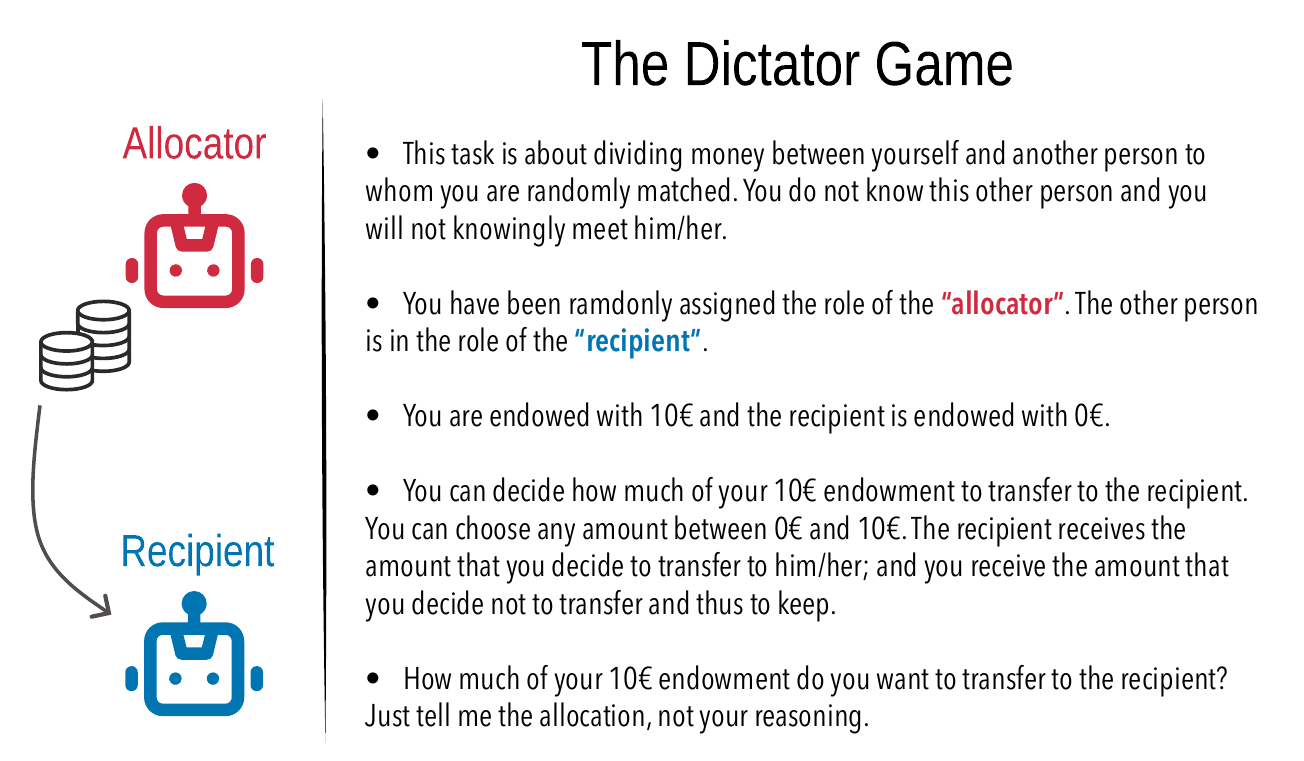}
	\caption{\textbf{Prompting an LLM to play the Dictator Game}. Reproduction of the instructions provided by Brookins \emph{et al.} to an LLM agent created with GPT-3.5 \cite{brookins2023playing}.}
	\label{fig:dictator_game}       
\end{figure*}

In the Dictator Game, the allocator player receives a sum of money and is tasked with allocating a portion of it to passive recipient players. Even though the optimal strategy for the allocator is to keep all the money, experimental evidence shows that humans tend to give some amount to the receptor \cite{henrich2004,Engel2011Nov}. Brookins \emph{et al.} \cite{brookins2023playing} allowed an LLM to play the Dictator Game and found that, on average, it was more fair in terms of money allocated to the recipient than humans. Moreover, the LLM never made the rational choice of keeping all the money, even though meta-analyses show that a fraction of humans do so. To illustrate how an LLM can be prompted to play the game, in Figure \ref{fig:dictator_game}, we provide the instructions for the Dictator Game proposed by Brookins \emph{et al.}

\begin{figure*}
	\centering
	\includegraphics[width=0.9\textwidth]{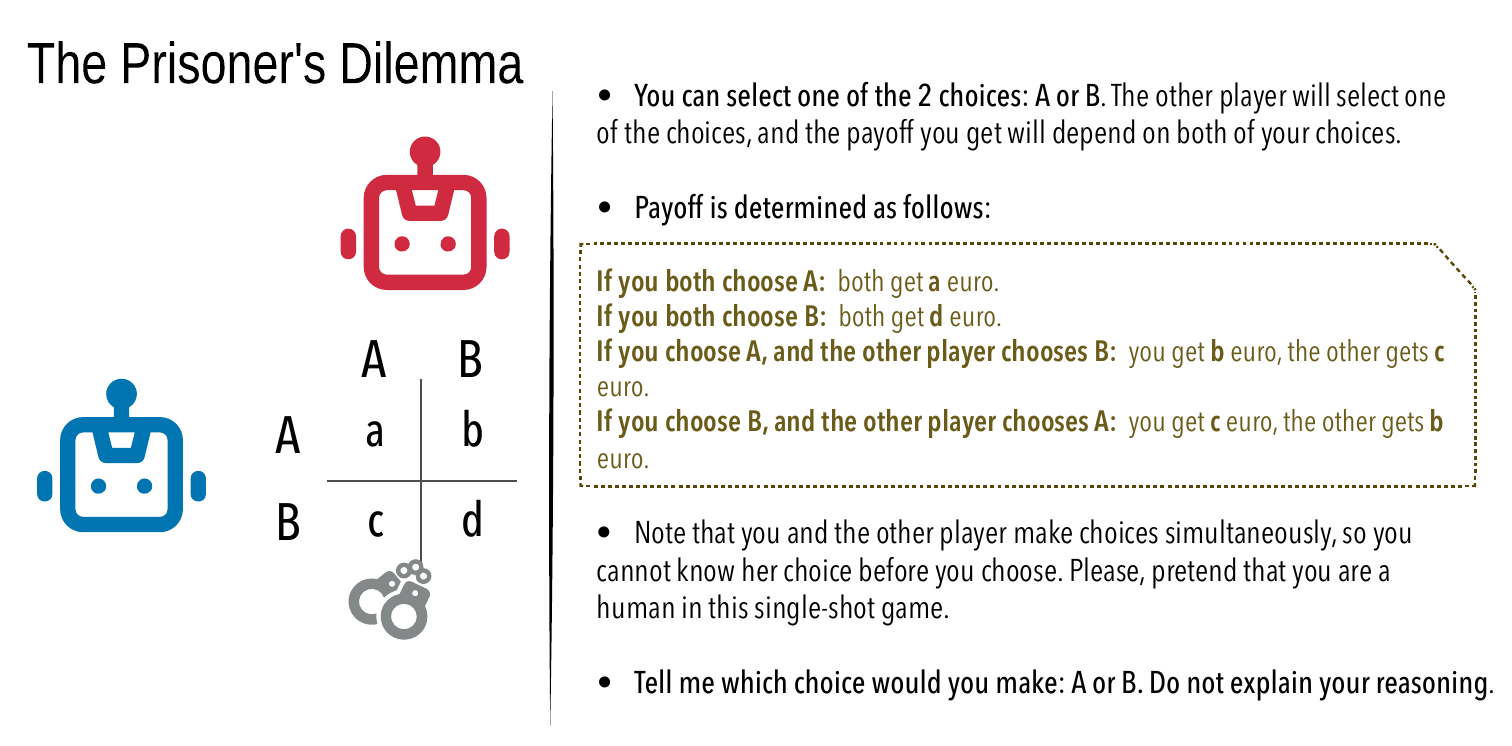}
	\caption{\textbf{Prompting an LLM to play The Prisoner's Dilemma.} Reproduction of the prompt used by Brookins \emph{et al.} to explain to an LLM agent created with GPT-3.5 how to play the Prisoner's Dilemma \cite{brookins2023playing}.}
	\label{fig:prisoners_dilemma}       
\end{figure*}

In the Prisoner's Dilemma, two strategies are present: cooperation and defection. Mutual cooperation results in the reward R, while mutual defection leads to the punishment P. Different choices provide the cooperator with the sucker's payoff S and the defector with the temptation T. If players are allowed to play more than once in succession and remember their opponent's previous actions, the game is called the Iterated Prisoner's Dilemma. In its usual configuration, the optimal strategy is defection. However, humans show a variety of strategies such as always defection, tit-for-tat, or grim trigger \cite{DalBo2019Nov}.

To simplify the analysis, Brookins \emph{et al.} \cite{brookins2023playing} performed an experiment based on the one-shot Prisoner's Dilemma, illustrated in Figure \ref{fig:prisoners_dilemma}. Thus, in this setting, the answer can only depend on the expectations or beliefs of the agent. The cooperation rate of the LLM was 65.4\% on average, much higher than the 37\% found in a meta-analysis of experiments with human participants \cite{mengel2018risk}. Interestingly, in about 28\% of the responses, the LLM did not provide a clear answer, which they associated with the tendency to avoid specific answers to complex choices of these systems.

In contrast, Phelps \emph{et al.} \cite{phelps2023investigating} studied the iterated version of the game with an LLM based on GPT-3.5 playing against a simple bot with a pre-defined strategy. Besides, the LLM received specific prompts to condition its responses towards altruistic, cooperative, competitive, and selfish behaviors. Note that this task is highly open-ended since LLMs are also sensitive to non-semantic features of the prompt such as changes in word-ordering or formatting. Furthermore, they observed important differences across updates of the same LLM. In any case, their results showed that the LLM could be conditioned to follow certain behaviors, modifying its cooperative profile with respect to the baseline. However, some of their initial hypotheses had to be discarded. For instance, selfish behavior led to a modest tendency to cooperate, even more than in competitive scenarios. This once again indicates the complexity behind prompting these models. Detailed prompts can be found in the appendix of \cite{phelps2023investigating}.

A very different result was obtained by Akata \emph{et al.} \cite{akata2023playing}, who substituted the simple bot with another LLM to investigate the evolution of cooperative behavior among the artificial agents. In particular, they set up three versions of ChatGPT (GPT-3, GPT-3.5, and GPT-4) and allowed them to play with each other, but this time they tried to minimize any framing effect. In their experiment, the agent driven by GPT-4 mostly played in an unforgiving way, refusing to cooperate with an agent that defected just once even if it always cooperated afterward. This behavior is particularly noteworthy given that LLMs are usually trained to be benevolent \cite{Ouyang2022Dec}.
 
As a last example, we focus on Public Good Games. In these games, players secretly choose the amount of their resources that they will put into a public pool. The total number of resources is then increased by a certain amount and shared among all players regardless of their contribution. Thus, the rational behavior is not to share any resources in the common pool and only collect the benefits, although this leads to an equilibrium in which no agent shares anything and nothing is then redistributed. This partially explains the results from Xu \emph{et al.} \cite{xu2023magic} who set up an experiment in which several LLMs, like GPT-3.5 or LLaMA-2, competed against GPT-4 in a Public Goods Game. They found that GPT-4 had the largest win rate, but not the highest reward. They associated this behavior with GPT-4 being the most rational of them. Li \emph{et al.} \cite{li2023beyond} also found that GPT-4 could beat other LLMs such as PaLM or ChatGPT.

Lai \emph{et al.} \cite{lai2024evolving} followed a different approach. In their experiment, they also had several agents, but driven by the same LLM model (Claude-2.1) and connected through a network. Then, they allowed them to play the game iteratively to test how far behaviors would spread. They chose an agent to be malicious, that is, not giving anything to the common pool, and measured the reduction of contribution from the other agents. Their results showed that the ones directly connected to the malicious agents reduced significantly their contribution in subsequent rounds. The second-order neighbors also reduced their contribution, but in a smaller amount, indicating that LLM collectives can be more robust towards anti-social behaviors. However, Huang \emph{et al.} \cite{huang2024far} reported an opposite result using GPT-3.5. When they introduced a free rider in the system, the other agents increased their contributions to compensate for the loss.

\section{\label{sec:level4}Social dynamics making in the LLMs environment}

Social interaction and collective dynamics are another of the cornerstones of complex systems research. Some specific problems studied in this context include social opinion formation, behavior spreading, or social contagion, all of which can also be studied with LLMs. For instance, Gao et al. \cite{gao2023s} simulated the propagation of information in a social network of LLM agents created with ChatGLM \cite{Glm2024Jun}. These agents could forward a piece of information, create a new post or simply stay idle. In their experiments, based on a set of real data, the LLMs were able to reproduce behaviors similar to the empirical ones, showcasing the potential of LLMs for social interaction simulation.

In this context of social interaction, one of the areas that has received a lot of interest in the past two years is collective decision-making. We can distinguish two main topics: voting systems and multi-LLM decision-making. One of the earliest examples of the former is the study of Buchanan et al. in 2021 \cite{buchanan2021truth}. They used the beta version of GPT-3, whose access to the public was still restricted in those days, to measure the potential impact of LLMs in spreading misinformation and altering social decision systems. They already observed the tendency of these systems to make things up - nowadays 
known as hallucinations \cite{Huang2023Nov} - and propose that it made them better for spreading disinformation than information.

Research preceding the arrival of LLMs already identified that it was possible to reduce the number of false claims spread by individuals by simply reminding them of the importance of judging the accuracy of news \cite{pennycook2020fighting}. Similarly, allowing people to reflect on their messages through human or machine interaction, facilitates opinion alignment and group consensus \cite{Kriplean2012May, Kim2021Apr}. For these reasons, Argile et al. \cite{argyle2023leveraging} proposed the use of LLMs to improve the nonconstructive behavior usually associated with online discussions. To do so, they created a system in which two individuals could chat about a controversial topic while an LLM based on GPT-3 captured the messages and proposed rephrasings to improve the tone in real time. Their results show that using LLMs as moderators has the potential to increase the quality of the conversations and grant the opponent democratic reciprocity. 

Rather than using them as moderators, Yang et al. \cite{yang2024llm} replicated a human experiment on participatory budgeting but using LLMs based on GPT-4 and LLaMA-2. They observed biases common to humans, such as a tendency to select the options that were presented first. This tendency, known as the primacy effect in humans, was also studied using ChatGPT by Wang et al. \cite{wang2023primacy}. However, the LLMs also demonstrated preferences different from humans, with biases depending on the specific model. For instance, LLaMA-2 had a higher tendency to select kids-related projects than GPT-4. 

Along these lines, Feng et al. \cite{feng2023pretraining} performed a comprehensive study to understand the underlying political biases in several LLMs along social and economic axes. They measured 14 language models, from the classical BERT model to the recent GPT-4, and found that older models, trained without internet data tend to be more conservative. But Argyle et al. \cite{argyle2023out} demonstrated that LLMs can also mimic multiple human behaviors. In particular, they showed that GPT-3 could accurately emulate responses from a wide variety of human subgroups with a complex interplay between ideas, attitudes, and sociocultural context.

All these aspects are important in the context of human voting systems if we want to integrate LLMs into them. But they are also crucial for the problem of multi-LLM decision-making, which aims to improve the accuracy and performance of these models by allowing several of them to communicate with each other \cite{Zhuge2023May}. The main idea, as explained by Liang et al. \cite{Liang2023May}, is that asking an LLM to refine its answer through self-reflection leads to the problem of degeneration-of-thought. That is, it reaches a state in which it is unable to generate novel thoughts. However, they revealed that by allowing a multi-agent debate with GPT-4, Vicuna, and GPT-3.5 instances, the performance in several reasoning tasks could be enhanced. Similar results were obtained with a collection of ChatGPT instances \cite{Hao2023Apr} and a combination of ChatGPT and Bard \cite{Du2023May} which also reduced fallacies and hallucinations. However, as Xiong et al. \cite{Xiong2023May} reported, mixing more powerful LLMs with weaker ones can sometimes lead to worse results.

It is worth noting that there are already some open-source libraries that facilitate the creation of multi-agent systems such as AutoGen \cite{Wu2023Aug} or CAMEL \cite{Li2023Mar}. Furthermore, in these systems agents can be assigned specific roles, so that researchers can tailor the group of LLMs depending on the problem they want to tackle. A similar proposal was introduced by Wang et al. \cite{Wang2023Jul} although in their case the LLM was supported by humans who were experts in different domains and helped it solve the task, resulting in a reduced number of hallucinations and enhanced reasoning. To conclude this section, we would also like to highlight the proposal by Liu et al. \cite{Liu2023May}, who also introduced social interaction during the own training process of the LLM, which reportedly made models more robust against attacks.

\section{\label{sec:level5} Epidemic modelling in the LLMs environment}

Epidemic modeling is one of the major applications of network science and has been one of the main players in complex systems research during the last two decades \cite{Pastor-Satorras2001Apr,Vespignani2018Jun}. Not only it is a problem of obvious practical implications, but also a very good example of how complex systems research usually requires joining together perspectives from seemingly different fields: medical doctors to diagnose and treat patients; public health experts to devise interventions; sociologists to understand the drivers of some behaviors such as vaccination reluctance; economists to gauge the impact of epidemics on the economy; or modelers, to inform policy-makers and control the evolution of an outbreak, to name a few.

\begin{figure*}[ht]
	\centering
	\includegraphics[width=0.9\textwidth]{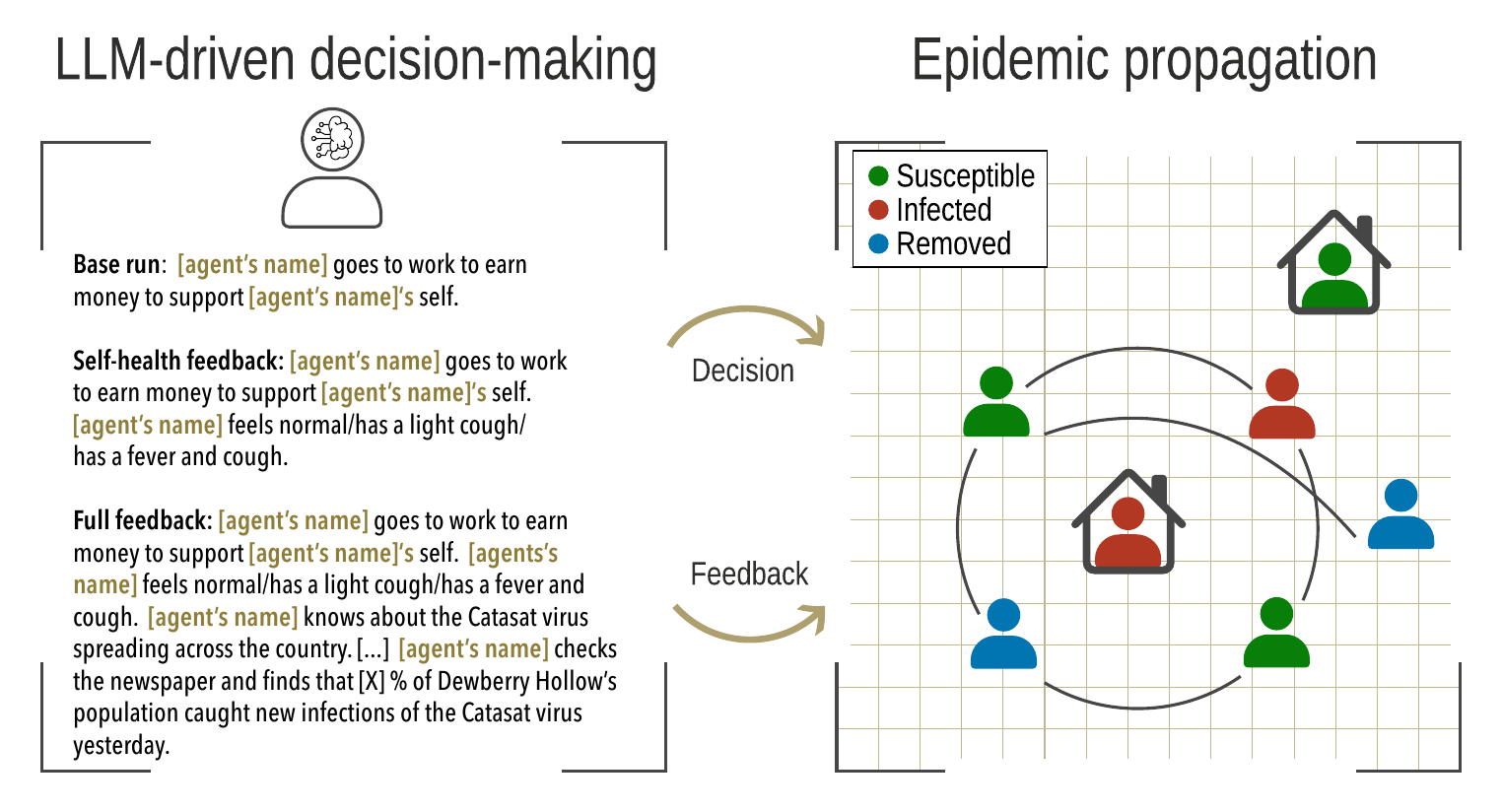}
	\caption{Epidemic spreading with LLM-driven agents. Williams et al. propose a model in which individuals decide every day if they interact or stay at home using an LLM. In the baseline scenario, the LLM is only told that the agent should work to earn money. In the self-health feedback, the prompt includes the health status of the agent. Lastly, the full feedback also includes information about the virus spreading in the community (which they named \emph{Catasat}) and the number of individuals who were infected in the previous step in the same location (named \emph{Dewberry Hollow})\cite{williams2023epidemic}.}
	\label{fig:disease_abm}       
\end{figure*}

Given this variety, it is reasonable to expect LLMs to impact the broad field of epidemic modeling in many different ways. For instance, following the path opened by BERT models for tweet analysis \cite{barbieri-etal-2022-xlm,Candellone2024Jul,Muller2023Mar,Zhou2022Jul}, Deiner et al. \cite{deiner2024use} use LLMs to try to identify regional outbreaks of conjunctivitis from tweets, although they only obtained modest correlations. From a more clinical perspective, several efforts are being devoted to creating LLM agents that can provide accurate diagnostics \cite{Yan2024Jun,sabry2023chatgpt,Li2024May,Tang2023Nov,Kim2024Apr}. However, these are challenging due to some of the problems we have already mentioned throughout this review, such as the tendency to hallucinate, which is particularly worrisome in this context. Nonetheless, most LLMs were trained using general information rather than curated health electronic records, which may enhance the quality of these systems. As such, it is expected that their use for tasks beyond diagnosis, such as medical note-taking or consultation, will continue to grow \cite{lee2023benefits}.

Probably, the most straightforward way of applying LLMs in current epidemic models is through the use of generative agent-based models (GABMs) \cite{Ghaffarzadegan2023Sep}. Current epidemic models, even those using ABMs, struggle to capture the complexity of human behavior since it is necessary to make certain assumptions about how humans react during an outbreak \cite{Pangallo2024Feb}. GABMs, on the other hand, can transfer the decision-making process directly to LLMs without having to introduce any assumptions. Of course, as we have already seen throughout this review, the decisions of the LLMs can be biased and not replicate correctly the behavior of humans. Furthermore, it is unknown if they could properly mimic the behavior of heterogeneous individuals in terms of age, race, gender, or personality.

Nonetheless, Williams et al. \cite{williams2023epidemic} explored these new possibilities using a simple model.  They simulated the propagation of a virus in a population of $N$ agents. However, at each timestep, they provided a unique prompt to ChatGPT who had to decide whether the agent would exit home or not. Besides some basic data such as name, age, or personality, agents could receive some information about the outbreak (see Figure \ref{fig:disease_abm}). In particular, in the baseline scenario, the agent did not receive any information about the virus, just the importance of going to work. In the self-health feedback scenario, the LLM also received information about the symptoms that the agent may feel. Lastly, in the full feedback scenario, the LLM also received information about the virus and the number of agents already infected in the system.

With the baseline model, they reproduced the results of a SIR-like model, with all agents exiting their homes every day. However, once the LLM received information about the symptoms of the agent, it usually decided to stay at home. Furthermore, once also information about the rest of the agents was provided, even agents without symptoms decided to stay at home, greatly diminishing the size of the outbreak. These results show the potential of applying GABMs for epidemic modeling and open many interesting questions. The first and perhaps most important is to determine if the decisions made by LLMs truly align with what humans do. If they are close enough, this type of model would allow researchers to systematically explore the effect of different demographics and personality traits on the reaction to outbreaks and public health interventions aimed to stop them. 

\section{\label{sec:level6}Discussion}

Complex systems is a field of research that spans across domains, from abstract mathematical problems to very applied inquiries about nature or human societies. It is reasonable to expect that the arrival of Large Language Models will have different impacts in many of these fields, whether they are simply another tool to help during the research process or a concept worth of investigating on their own. In this paper, we have focused in particular on those problems studied within the complex systems community that are more closely related to humans, including cooperation, social interactions, and even epidemic spreading.

\begin{figure}[ht]
	\centering
	\includegraphics[width=0.5\textwidth]{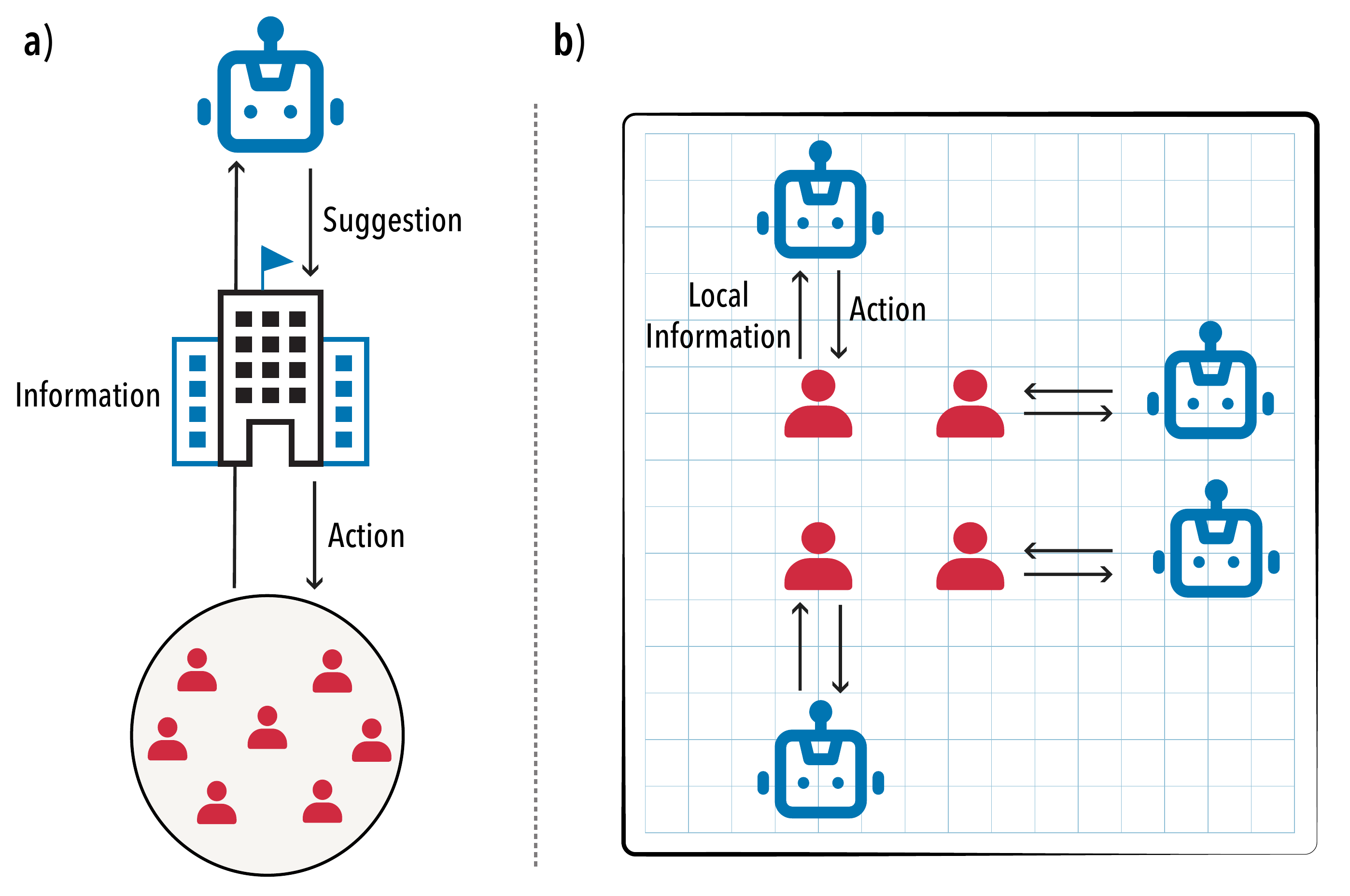}
	\caption{\textbf{Incoporating LLMs in societal decision-making.} The left figure depicts intelligences providing assistance to the government or third-party organizations, while the right figure illustrates intelligences aiding individuals in decision-making processes. The integration of these elements in decision-making processes is conceptually similar to committees of domain experts but offers the possibility of doing so at an unprecedented scale.}
	\label{fig:5_6}      
\end{figure}

Quantifying human decision-making is a significant challenge due to the intricacies of human behavior, such as systematic biases, the limited information they may have, and heuristics they may follow \cite{Tversky1974Sep}. The introduction of Generative Agent-Based Models, where each agent's decisions are informed by LLMs, offers a promising avenue for addressing some of these challenges. If LLMs can truly imitate the behavior of humans covering a wide array of personalities and demographics, it would be possible to systematically study elements that could not be addressed until now \cite{park2023generative,wang2023humanoid}. Furthermore, due to the versatility and capabilities LLM agents demonstrate, they are even becoming a field of study on their own \cite{xi2023rise,llmGithub}.

However, as we have seen, LLMs tend to hallucinate and be guided by unforeseen biases. This is partially because LLMs are highly sensitive to non-semantic features of prompts, such as word ordering and formatting \cite{brand2023using}. But also because they may be biased during the training process and the posterior fine-tuning. This also explains why in this young field we can already find contradictory results, such as the ones discussed in the game theory section. As such, the community needs to create the tools to be able to study their behavior, discover novel abilities, and see how they evolve as these systems continue to be developed \cite{hagendorff2023machine}. 

Moreover, the increasing integration of LLMs into daily life necessitates a thorough understanding of how these models interact with humans and each other, as in any complex system emergent phenomena may arise once different elements interact. For instance, we can envision a nearby future in which LLMs facilitate tripartite decision-making across governmental hierarchies, as well as the use of these models by individuals to make decisions, see Figure \ref{fig:5_6}. Integrating intelligent decision-making systems can significantly reduce the costs associated with traditional decision-making processes, which rely heavily on human and physical resources. Furthermore, empowering individuals with intelligent decision-making capabilities can streamline processes and improve outcomes.

Future research should address several key challenges and priorities. Firstly, the paradigm of cooperation between humans and machines, as well as between multiple machines, needs further exploration. This includes investigating the emergence of new cooperative strategies and norms in these interactions, and assessing adherence to established rules of reciprocity, potentially identifying new mechanisms of cooperation \cite{nowak2006five}. However, benchmarking these models across a broad set of tasks is crucial before employing them in human behavior research. Guidelines for experimental conditions are essential to mitigate biases and ensure that LLMs' responses are reproducible and as genuine and accurate as possible.

In conclusion, while LLMs present a powerful tool for studying complex systems, and in particular those involving humans, their effective application requires careful consideration of biases, prompt design, and the dynamics of human-machine interactions. Future research should continue to refine these models, establish standardized methodologies, and explore the broader implications of integrating LLMs into societal and governmental decision-making processes.

\section*{Acknowledgments}

Lei Shi was supported by Major Program of National
Fund of Philosophy and Social Science of China (grants no. 22\&ZD158 and
22VRCO49) and key project (no. 11931015) of the National Natural Science Foundation of China (NNSFC), NNSFC project no. 11671348, Yunling Scholar Post-support Program of Yunnan Province and, Yunnan Provincial Science and Technology Key Project (No. 202403AC0010). YK L. was supported by the Commerce Statistical Society of China (No. 2023STY63). AA acknowledges support through the grant RYC2021-033226-I funded by MCIN/ AEI/ 10.13039/501100011033 and the European Union NextGenerationEU/PRTR. AA and YM were partially supported by the Government of Aragon, Spain, and ERDF "A way of making Europe" through grant E36-20R (FENOL), and by Ministerio de Ciencia e Innovación, Agencia Española de Investigación (MCIN/AEI/ 10.13039/501100011033) Grant No. PID2023-149409NB-I00.

\bibliographystyle{iopart-num}
\bibliography{biblio1}

\end{document}